\input{aipcheck}

\documentclass[
    ,final            
  ]
  {aipproc}

\layoutstyle{8x11double}

\begin{document}

\title{Puzzles in quarkonium hadronic transitions with two pion emission}


\author{F. Fern\'andez}{
  address={Grupo de F\'isica Nuclear and IUFFyM, Universidad de Salamanca, E-37008 Salamanca, Spain}
}

\author{J. Segovia}{
  address={Grupo de F\'isica Nuclear and IUFFyM, Universidad de Salamanca, E-37008 Salamanca, Spain}}

\author{P.G. Ortega}{
  address={CERN, CH-1211 Geneva, Switzerland}
  }

\author{D.R. Entem}{
    address={Grupo de F\'isica Nuclear and IUFFyM, Universidad de Salamanca, E-37008 Salamanca, Spain}}

\begin{abstract}
 The anomalously large rates of some hadronic transitions from quarkonium are studied using QCD multipole expansion (QCDME) in
the framework of a constituent quark model which has been successful in describing hadronic
phenomenology. The hybrid intermediate states needed in the QCDME method are calculated
in a natural extension of our constituent quark model based on the Quark Confining
String (QCS) scheme. Some of the anomalies are explained due to the presence of an
hybrid state with a mass near the mass of the decaying resonance whereas other are
justified by the presence of molecular components in the wave function. Some 
unexpected results are pointed out.
\end{abstract}

\maketitle


\section{Introduction}

The recent experimental data of the hadronic transitions of heavy quarkonia, such as $\psi 
(nS)$ or $\Upsilon(nS)$ to lower states with emission of two pions, show a puzzling behavior. 
The experimental data show that the $X(4260)$, $X(4360)$ and $X(4660)$
states apparently decay only through particular channels. The $X(4260)$ resonance
has been seen only in $J/\psi\pi^{+}\pi^{-}$, whereas the $X(4360)$ and $X(4660)$ appear
only in the $\psi(2S)\pi^{+}\pi^{-}$ channel. Furthermore, these two resonances have
an anomalously large width~\cite{Lees:2012cn,Wang:2007ea,Lees:2012pv}. Moreover, such
decays have not been observed for the $\psi(4415)$. In the bottom sector, compared to
the ordinary $\Upsilon(nS)\rightarrow \Upsilon(mS)$ $(m<n)$ transitions, the partial width of the
$\Upsilon(10860)$ is out of line by two orders of magnitude~\cite{Abe:2007tk}.

The typical momentum involved in the transition is too low to use perturbative QCD and therefore such methods can not be applied. Then, non perturbative methods, like the QCD multipole expansion approach  (QCDME), should  be used to describe at least the transition of the lower lying states. In the single channel picture of QCDME the heavy quarkonium system serves as a compact color source which
emits two soft gluons which hadronize into two pions. After the emission of the first
gluon and before the emission of the second one there exists an intermediate state where
the $Q\bar{Q}$ pair together with the gluon forms an hybrid state. The width of the
transition critically depends on the particular spectrum of the hybrid states, therefore it is
important to describe the $Q\bar{Q}$ states an the hybrid consistently using as few
parameters as possible. 

Apart from lattice
calculations~\cite{Juge:1999ie,Dudek:2008sz}, hybrid meson properties have been
calculated in different models: the flux-tube
model~\cite{Isgur:1984bm,Barnes:1995hc}, constituent gluons~\cite{Horn:1977rq},
Coulomb gauge QCD~\cite{Guo:2008yz} and quark confining string model
(QCS)~\cite{Tye:1975fz,Giles:1977mp,Buchmuller:1979gy} or QCD string
model~\cite{Kalashnikova:2008qr}. Among them we adopt
the QCS model since it was used in the early works of QCDME and it incorporates
finite quark mass corrections.

Above the open flavor threshold, the QCDME scenario may change due to the possible contribution of molecular components~\cite{JPG1},
which can modified the hadronic decay width.

In this work we will address the description of the new data of the hadronic
transitions in heavy quarkonium using the QCDME in the framework of a constituent quark model (see
references~\cite{Valcarce:2005em} and~\cite{Segovia:2013wma} for reviews) which has
been successful in describing the hadron phenomenology and the hadronic reactions.
Hybrid states are consistently generated in the original quark model using the QCS
scheme. In this way we minimize the number of free parameters. Above the open flavor threshold molecular meson-meson channels coupled to the $c\bar c$ states are included in the calculation

\section{Theoretical framework}
The constituent quark model we use is based on the assumption that the spontaneous chiral symmetry breaking  generates the constituent quark mass. To compensate this mass term in the Hamiltonian the Lagrangian must include Goldstone boson fields which mediates the interaction between quarks. The minimal realization of this mechanism include a pseudoscalar boson and an scalar one.
 This fact does not affect the heavy quark sector but is of paramount importance in the molecular picture because the only remaining  interaction between the two molecular component, due to its color singlet nature, is the one driven by the Goldstone boson exchanges between the light quarks. Below the chiral symmetry scale quarks still interact through one gluon exchange and confinement potential. Explicit expression for this interactions are given in Ref.~\cite{Vij}.

In this model mesons are described as clusters of one pair of quark and antiquark.
To found the quark-antiquark bound states  we solve the 
Schr\"odinger equation using the Gaussian Expansion Method~\cite{r20}. In the case of the molecular structures the four body problem is  also solved using the gaussian expansion and the  two body wave functions obtained from the solution of the  Schr\"odinger equation.  To derive the meson-meson interaction from the $qq$ interaction we use the Resonating Group Method (RGM)~\cite{r22}.  The coupling between the $q\bar q$ and the four quarks configuration are performed using the $^3P_0$ model~\cite{r21}. A more detailed discussion of the model and its application to the states mentioned above can be found in Ref.~\cite{JPG1}.

To calculate the hadronic transition we use the QCDME  method. The multipole expansion has been widely used for studying radiation processes in which the electromagnetic field is radiated from local sources. In our case the energy difference between the initial and the final states is usually small and therefore the gluon wavelength is large compared to the typical size of the $Q\bar Q$ states and can be treated in a multipole expansion.  The gauge-invariant formulation of multipole expansion within QCD was given by T.-M. Yan in Ref.~\cite{Yan}. 

The two pion hadronic transitions are dominated by double electric-dipole transitions (E1E1). The transition amplitude  split into two factors. The first one concerns to the wave functions 
and energies of the initial and final quarkonium state as well as those of the intermediate states. All these quantities can be calculated using suitable quark models. The second one describes the conversion of the emitted gluons into light hadrons. Its scale is the light hadron mass, which is very low and therefore  cannot be calculated in a perturbative way.  Usually one uses a phenomenological approach based on PCAC ~\cite{r23} which involves two parameters ($C_1$ and $C_2$). The $C_1$ term is isotropic (S-wave) while the $C_2$ term is angular dependent (D-wave). These two parameters are fitted to the
well established $\psi(2S)\to J/\psi\pi^{+}\pi^{-}$ and $\psi(3770)\to J/\psi \pi^{+}
\pi^{-}$ transitions.

As explained above, these transitions involve the emission of two gluons that hadronized into  two pions.  The intermediate states after the emission
of the first gluon and before the emission of the second gluon are states with
a gluon and a  $Q\bar{Q}$ pair which are the so-called hybrid states. We describe these states in the QCS model.

The QCS model is defined by a relativistic-, gauge- and reparametrization-invariant
action describing quarks interacting with color $SU(3)$ gauge fields in a two
dimensional world sheet. The model has no gluonic degrees of freedom, but has instead
string degrees of freedom.

The string can carry energy-momentum only in the region between the quark and the
antiquark, thus the quarks appear to be at the ends of the string. The equation that  describes the dynamics of the quark-antiquark pair linked  by the string is the usual  Schr\"odinger
equation with a confinement potential. Gluon excitation effects are described by the
vibration of the string. These vibrational modes provide new states beyond the naive
meson picture. A complete description of the QCS model can be found in Ref.~\cite{Giles:1977mp}. Its generalization to our constituent quark model is described in Ref.~\cite{seg2}.

\section{Results}

All the parameters of the quark model are taken from Ref.~\cite{seg}. The quantum numbers of hybrid
states which participate in the two pion transition are $J^{PC}=1^{--}$. The mass of the two low lying  states in the $c\bar{c}g$ sector are  $4.35$ and $4.64\,{\rm GeV}$. The first one
agrees with the results of the flux-tube model~\cite{Barnes:1995hc} $(4.1-4.2)$,
Coulomb gauge QCD~\cite{Guo:2008yz} $(4.47)$, QCD string
model~\cite{Kalashnikova:2008qr} $(4.40)$, potential models ~\cite{HWKE}
$(4.23)$ and lattice
calculation~\cite{Dudek:2008sz} $(4.40)$. In the bottom sector the ground state mass is $10.785\,{\rm GeV}$ near to the mass of the potential model $(10.79)$ ~\cite{HWKE}

\begin{table}[!t]
\begin{tabular}{c|cc|cc}
\hline
\hline
Initial Meson & $R^{\rm th}_{J/\psi}$ (eV) & $R^{\rm ex}_{J/\psi}$ (eV) &
$R^{\rm th}_{\psi(2S)}$ (eV) & $R^{\rm ex}_{\psi(2S)}$ (eV) \\
\hline
$\psi(4040)$ & $1.20$ & - & 0.11 & - \\
$\psi(4160)$ & $0.40$ & - & $6\times 10^{-2}$ & -\\
$X(4360)$    & $52.5$ & - & 5.05 & $7.4\pm 0.9$ \\
$X(4415)$    & $3\times 10^{-6}$ & - & 0,27 & - \\
$X(4660)$    & $0.58$ & - & 1.08 & $1.04\pm 0.5$ \\
\hline
\hline
\end{tabular}
\caption{\label{tab:pipicc} $ R_{\psi(nS)} = {\cal B}_{\pi^{+}\pi^{-}\psi(nS)} \times
\Gamma_{e^{+}e^{-}}$ for the $J^{PC}= 1^{--}$ charmonium states. Experimental data
are from Ref.~\cite{Lees:2012pv}.}
\end{table}

\begin{table}[!t]
\begin{tabular}{c|cc|cc}
\hline
\hline
Initial Meson & $\sigma^{\rm th}_{J/\psi}$ (pb) & $\sigma^{\rm ex}_{J/\psi}$ (pb) &
$\sigma^{\rm th}_{\psi(2S)}$ (pb) & $\sigma^{\rm ex}_{\psi(2S)}$ (pb) \\
\hline
$\psi(4040)$ & $13.46$           & - & $1.25$  & -         \\
$\psi(4160)$ & $3.32$            & - & $0.50$  & -         \\
$X(4360)$    & $329.7$           & - & $31.69$ & $52\pm 2$ \\
$X(4415)$    & $4\times 10^{-5}$ & - & $3.38$  & -         \\
$X(4660)$    & $10.97$           & - & $20.29$ & $28\pm 2$ \\
\hline
\hline
\end{tabular}
\caption{\label{tab:cscc} The cross section at peak for the $J^{PC}= 1^{--}$ $S$-wave
charmonium states. Experimental data are from Ref.~\cite{Lees:2012pv}.}
\end{table}

\begin{table}[!t]
\begin{tabular}{c|cc|cc|cc}
\hline
\hline
Initial Meson & $R^{\rm th}_{\Upsilon (1S)}$ (eV) & $R^{\rm ex}_{\Upsilon(1S)}$ (eV)
& $R^{\rm th}_{\Upsilon(2S)}$ (eV) & $R^{\rm ex}_{\Upsilon(2S)}$ (eV) &
$R^{\rm th}_{\Upsilon(3S)}$ (eV) & $R^{\rm ex}_{\Upsilon(3S)}$
(eV) \\
\hline
$\Upsilon(2S)$    & $98.34$ & $105.4\pm 4.3$ & -    & - & - & -\\
$\Upsilon(3S)$    & $23.94$ & $18.5\pm 9.8$  & 5.58 & -\\
$\Upsilon(4S)$    & $6\times 10^{-2}$        & $(2.3\pm 0.9)\times 10^{-2}$ & $2.5\times10^{-3}$  & $(2.3\pm 0.4)\times 10^{-2}$   & - & -\\
$\Upsilon(10860)$ & $4.1\times 10^{-2}$      & $1.64 \pm 0.40$ & $5.8 \times 10^{-2}$ & $ 2.42\pm 0.64$ & $1.8\times 10^{-2} $ & $1.49\pm 0.65$ \\
\hline
\hline
\end{tabular}
\caption{\label{tab:pipibb} $R_{\Upsilon(nS)} = {\cal B}_{\pi^{+}\pi^{-}\Upsilon(nS)}
\times \Gamma_{e^{+}e^{-}}$ for the $J^{PC}= 1^{--}$ bottomonium states. Experimental
data are  from Ref.~\cite{Olive}.}
\end{table}

Table~\ref{tab:pipicc} shows the calculated ${\cal B}_{\pi^{+}\pi^{-}\psi(nS)} \times
\Gamma_{e^{+}e^{-}}$ for the $J^{PC}= 1^{--}$ charmonium states. As the decays
$\psi(2S)\to J/\psi \pi^{+}\pi^{-}$ and $\psi(3770)\to J/\psi \pi^{+}\pi^{-}$ have
been used to fit the $C_{1}$ and $C_{2}$ parameters, they are not included in the
table. One can see that in the case of the decay channel $\psi(2S)\pi^{+}\pi^{-}$ the
only significant values correspond to the decays of the $X(4360)$ and $X(4660)$
which are also in agreement with the recent experimental data. 

 In the decay channel $J/\psi\pi^{+}\pi^{-}$ a high value of the
${\cal B}_{\pi^{+}\pi^{-}\psi(nS)}\times\Gamma_{e^{+}e^{-}}$ is obtained for the
$X(4360)$ resonance. This result apparently contradicts the experimental data because
this decay has not be reported in the reaction $e^{+}e^{-}\to
J/\psi\pi^{+}\pi^{-}$~\cite{Lees:2012cn}. The cross section of this reaction shows a
resonance in the $4.2-4.4$ GeV. energy region which has been attributed to the $X(4260)$.
This resonance does not appear in our calculation as a $c\bar{c}$ meson and its nature is still under discussion. An
interference between the $X(4260)$ and $X(4360)$ resonances would be possible and should be explored. 

If one looks to the values of the cross section at peak (Table~\ref{tab:cscc}) one sees that in the decay channel $J/\psi\pi^{+}\pi^{-}$  the values of the rest of peak cross sections, besides the resonance at $4.36$ GeV, are of the order
of the experimental background and it is difficult to decide
whether a resonance is present or not. These results justify
why no signal of the $\psi(4040)$, $\psi(4160)$ and $\psi(4415)$ has been reported.

The results for the bottomonium sector are shown in Table~\ref{tab:pipibb}. One can see that the theoretical values agree reasonably well 
with the experimental ones except in the case of the  $\Upsilon(10860)$. 
In this case we do not find any hybrid state near its energy  and the
mechanism which explains the large width in the charm sector cannot be applied to this
case.

The  $\Upsilon(10860)$ is above the open-bottom threshold  and then contributions of molecular components of the wave function like $BB$, $BB^*$ or  $B^*B^*$ may appear. The importance of these components lies in the fact that the decays of the molecular structures through the $\Upsilon \pi \pi$ channel is OZI allowed whereas the decay of a $b\bar b$ component is OZI forbidden. Therefore the presence of molecular components in the wave function enhances the transition probability. 
\begin{table}[!t]
\begin{tabular}{ccccc}
\hline
\hline
Mass (MeV) & Width (MeV). & Prob $(5^3S_1)$ & Prob $(4^3D_1)$ & Prob(Mol) \\
\hline
$10829.10$    & $75.31$ & $0.72$ & 11.27 & 88.01 \\
$10849.73$    & $15.51$ & $0.24$ & 2.38 & 97.38 \\
$10849.93$    & $89.43$ & $ 3.48$ & 11.99 & 84.53 \\

\hline
\hline
\end{tabular}
\caption{\label{tab:molecule} Molecular components of the $1^{--}$ bottomonium states above the $BB$ threshold..}
\end{table}

\begin{table}[!t]
\begin{tabular}{cccc}
\hline
\hline
Channel & Width (MeV). & B.R.[$\%$] & B.R. exp [$\%$]  \\
\hline
$\Upsilon (1S)\pi \pi$    & $0.330$ & $0.362$ & $0.53\pm 0.06$ \\
$\Upsilon (2S)\pi \pi$    & $0.023$ & $0.026$ & $0.78\pm 0.13$  \\
$\Upsilon (3S)\pi \pi$    & $0.004$ & $0.004$ & $0.48\pm 0.19$  \\

\hline
\hline
\end{tabular}
\caption{\label{tab:BR} Widths and Branching Ratios of the $\Upsilon (nS)$ hadronic transitions through molecular components. Experimental
data are from Ref.~\cite{Olive}.}

\end{table}

\begin{table}[!t]
\begin{tabular}{ccccc}
\hline
\hline
Channel & $f_0(500)$ & $f_0(960)$ & Nonresonant  \\
\hline
$\Upsilon (1S)\pi \pi$    &  $5.27\times 10^{-4}$ & $1.14\times 10^{-3}$ &$ 0.330$ \\
$\Upsilon (2S)\pi \pi$    &  $9.22\times 10^{-4}$ & $1.30\times 10^{-2}$ & $1.03\times 10^{-2}$ \\
$\Upsilon (3S)\pi \pi$    &  $5.97\times 10^{-4}$ & $2.86\times 10^{-3}$ & $1.98\times 10^{-4}$ \\

\hline
\hline
\end{tabular}
\caption{\label{tab:BR_Res} Resonant and Nonresonant contribution to the $\Upsilon(nS)$ hadronic transition .}

\end{table}

Following Ref.~\cite{JPG1}, we have performed a coupled channel calculation including the $b \bar b$ $5^3S_1$ and $4^3D_1$ together with the $BB$, $BB^*$ and  $B^*B^*$ molecules. The result of this calculation is shown in Table~\ref{tab:molecule}. We obtain two resonances in the $10860$ MeV mass region. The first one is too narrow and  the second is wide. We identify the wide resonance with the $\Upsilon(10860)$. The process $\Upsilon(10860)\rightarrow \Upsilon(nS) \pi\pi$ involves the creation of a $q\bar q$ pair and a quark rearrangement process. This can result into a nonresonant final state $\Upsilon(nS) \pi\pi$, or an intermediate state in which the two pion forms a resonance (like $f_0(980)$) following by a subsequent decay into the final channel. 
In the calculation we include the $f_0(500)$ and the $f_0(980)$ resonances together with the nonresonant contribution. Results are shown in Tables~\ref{tab:BR} and~\ref{tab:BR_Res}. The contribution of the molecular components is able to explain the branching ratio of the $\Upsilon(10860)\rightarrow\Upsilon (1S)\pi \pi$ but fails in the $\Upsilon (2S)\pi \pi$ and $\Upsilon (3S)\pi \pi$ cases. In all cases the main contribution corresponds to the nonresonant process but this contribution decreases as the phase space decreases. The contribution of more resonances could help to improve the results. 

\vspace*{-0.20cm}
\section{Summary}
Using the constituent quark model quoted above we calculate the $J/\psi \pi \pi$ and $\psi(2S) \pi \pi$ decays of the $J^{PC}=1^{--}$ charmonium and bottomonium states. Hybrid states are consistently generated in the original quark model using the quark confining string (QCS) scheme. Above threshold we incorporate the effects of molecular components.  
We are able to explain the anomalously large decay width of the $X(4360)$ and $X(4660)$ due to the presence of hybrid states located near the masses of both resonances. However this mechanism does not work in the case of the $\Upsilon(10860)$. The contribution of molecular components improves the description of the $\Upsilon(10860)\rightarrow\Upsilon (1S)\pi \pi$ branching ratio but fails to explain the branching ratio in the $\Upsilon (2S)\pi \pi$ and $\Upsilon (3S)\pi \pi$ cases. 
Our model also predicts a large value for the decay $X(4360)\rightarrow J/\psi \pi \pi$ which has been not seen in the experiments.  Therefore the puzzling situation of the hadronic decays still deserves more experimental and theoretical efforts.

\vspace*{-0.20cm}
\section{Acknowledgements}

This work has been partially funded by U.\,S.\,Department of Energy, Office of
Nuclear Physics, contract no.~DE-AC02-06CH11357, by Ministerio de Ciencia y
Tecnolog\'\i a under Contract no. FPA2010-21750-C02-02, by the European
Community-Research Infrastructure Integrating Activity ``Study of Strongly
Interacting Matter'' (HadronPhysics3 Grant no. 283286) and by the Spanish
Ingenio-Consolider 2010 Program CPAN (CSD2007-00042).



\bibliographystyle{aipproc}   





\end{document}